\documentstyle[prl,aps,psfig]{revtex} 
\begin{document}
\twocolumn[\hsize\textwidth\columnwidth\hsize\csname
@twocolumnfalse\endcsname 
\title{The neutron radii of $^{208}$Pb and neutron stars}
\author{C.~J.~Horowitz}
\address{Nuclear Theory Center and Dept. of Physics, Indiana
University, Bloomington, IN 47405}
\author{J.  Piekarewicz}
\address{Department of Physics Florida State University, 
Tallahassee, FL 32306}
\date{\today} 
\maketitle 
\begin{abstract}
A new relation between the neutron skin of a heavy nucleus and the
radius of a neutron star is proposed: the larger the neutron skin of
the nucleus the larger the radius of the star.  Relativistic models
that reproduce a variety of ground-state observables can not determine
uniquely the neutron skin of a heavy nucleus. Thus, a large range of
neutron skins is generated by supplementing the models with nonlinear
couplings between isoscalar and isovector mesons. We illustrate how
the correlation between the neutron skin and the radius of the star
can be used to place important constraints on the equation of state
and how it may help elucidate the existence of a phase transition in
the interior of the neutron star.
\end{abstract}
\vskip2.0pc]

What determines the size of a neutron star? For spherical, static
stars in hydrostatic equilibrium, the so-called Schwarszchild stars,
the sole feature responsible for their size is the equation of state
of neutron-rich matter. The skin of a heavy nucleus --- a system 18
orders of magnitude smaller and 55 orders of magnitudes lighter than a
neutron star --- is also composed of neutron-rich matter, although at
a lower density.

In a recent publication we studied the relation between the neutron
skin of ${}^{208}$Pb and the non-uniform solid crust of a neutron
star~\cite{Ho01}. For models with a stiff equation of state it is 
energetically unfavorable to separate uniform nuclear matter into 
regions of high and low densities. Thus models with a stiff equation 
of state predict low transition densities from non-uniform to uniform 
neutron-rich matter and consequently thinner crusts. The thickness 
of the neutron skin in ${}^{208}$Pb also depends on the equation of 
state of neutron-rich matter. The stiffer the equation of state the
thicker the neutron skin. Thus, an inverse relationship was
established: the thicker the neutron skin of a heavy nucleus the
lower the transition from non-uniform to uniform neutron-rich matter.

In this work we study the relation between the neutron skin of a heavy
nucleus and the radius of a neutron star. Indeed, we will show that
models with thicker neutron skins produce neutron stars with larger
radii. Such a study is particularly timely as it complements important
advances in both experimental physics and observational
astronomy. Indeed, a proposal now exists at the Jefferson Laboratory
to measure the neutron radius of $^{208}$Pb via parity-violating
electron scattering~\cite{prex,bigpaper}. Moreover, a number of
improved radii-measurements on isolated neutron stars, such as
Geminga~\cite{Go99}, RX J185635-3754~\cite{Wa97,Wa01,Po01},
Vela~\cite{Li99,Pa01}, and CXOU 132619.7-472910.8~\cite{Ru01} are now
available. While these measurements are not yet accurate enough to set
stringent limits on the equation of state, they represent an important
first step in that direction~\cite{Ha01}.

Our starting point will be the relativistic effective-field theory
of Ref.~\cite{horst} supplemented with new couplings between the
isoscalar and the isovector mesons. The interacting Lagrangian 
density for this model is given by~\cite{Ho01,horst}
\begin{eqnarray}
{\cal L}_{\rm int} &=&
\bar\psi \left[g_{\rm s}\phi   \!-\! 
         \left(g_{\rm v}V_\mu  \!+\!
    \frac{g_{\rho}}{2}{\mbox{\boldmath $\tau$}}\cdot{\bf b}_{\mu} 
                               \!+\!    
    \frac{e}{2}(1\!+\!\tau_{3})A_{\mu}\right)\gamma^{\mu}
         \right]\psi \nonumber \\
                   &-& 
    \frac{\kappa}{3!} (g_{\rm s}\phi)^3 \!-\!
    \frac{\lambda}{4!}(g_{\rm s}\phi)^4 \!+\!
    \frac{\zeta}{4!}   g_{\rm v}^4(V_{\mu}V^\mu)^2 
    \nonumber \\
                   &+&
    g_{\rho}^{2}\,{\bf b}_{\mu}\cdot{\bf b}^{\mu}
    \left[\Lambda_{\rm s} g_{\rm s}^{2}\phi^2 +
          \Lambda_{\rm v} g_{\rm v}^{2}V_{\mu}V^\mu\right] \;.
 \label{LDensity}
\end{eqnarray}
The model contains an isodoublet nucleon field ($\psi$) interacting
via the exchange of two isoscalar mesons, the scalar sigma ($\phi$)
and the vector omega ($V^{\mu}$), one isovector meson, the rho (${\bf
b}^{\mu}$), and the photon ($A^{\mu}$). In addition to meson-nucleon
interactions the Lagrangian density includes scalar and vector
self-interactions. (Note that while the original model allows for
$\rho$-meson self-interactions~\cite{horst}, their phenomenological
impact has been documented to be small so they will not be considered
in this contribution). The scalar self-interaction is responsible for
reducing the compression modulus of nuclear matter from the
unrealistically large value of $K\!=\!545$~MeV~\cite{Wa74,Se86}, all
the way down to about $K\!=\!230$~MeV. This latter value appears to
be consistent with the isoscalar giant-monopole resonance (GMR) in
${}^{208}$Pb~\cite{Yo99,Pi00}. Omega-meson self-interactions have
proven essential for the softening of the equation of state at high
density. Indeed, without them large limiting masses for neutron stars
(of about $2.8 M_{\odot}$) are predicted, even for the softer models
that provide a good description of the giant-monopole
resonance~\cite{Pi00}. This is because the GMR constrains the equation
of state around saturation density but leaves the high-density
behavior practically undetermined. Models that include omega-meson
self-interactions soften the high-density equation of state to such 
a degree that limiting masses of $1.8 M_{\odot}$ become 
possible~\cite{horst}. Finally, the nonlinear couplings 
$\Lambda_{\rm s}$ and $\Lambda_{\rm v}$ are included to modify 
the density-dependence of the symmetry energy~\cite{Ho01}.

We compute the neutron radius of ${}^{208}$Pb and the radius of a
``canonical''1.4 solar-mass neutron star for all the parameter
sets listed in Ref.~\cite{Ho01}. One of these is the very 
successful NL3-model of Lalazissis, K\"onig, and Ring.~\cite{NL3}. 
The other models (S271 and Z271) were introduced in Ref.~\cite{Ho01}. 
All the sets have been constrained to reproduce three important 
properties of symmetric nuclear matter at saturation: the saturation 
density ($1.3\,{\rm fm}^{-1}$), the binding-energy per nucleon 
($-16.25$~MeV), and the compression modulus ($271$~MeV). The value 
of the effective nucleon mass at saturation, which is not accurately 
known, as well as the strength of the omega-meson self-coupling 
coupling ($\zeta$) differ in the various models~\cite{Ho01}. This 
in turn permits a modification of the high-density component of 
the equation of state in the different models.

\begin{figure}[h]
\leavevmode\centering\psfig{file=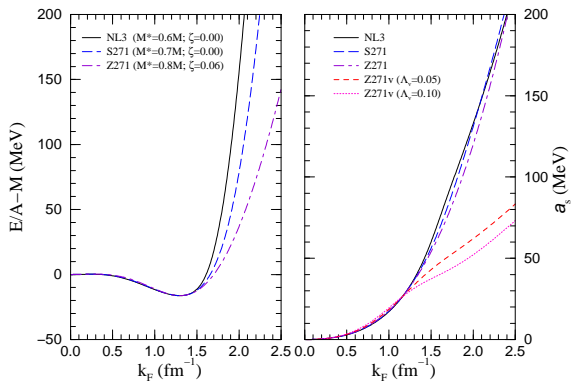,width=3in}
\caption{Binding-energy per nucleon in symmetric
         nuclear matter (left panel) and symmetry
	 energy (right  panel) in the various models
	 discussed in the text.}
 \label{Fig1}
\end{figure}

The energy density of symmetric nuclear matter can be computed in
a mean-field approximation by solving the classical equations of
motion for the meson fields. In the mean-field limit it is given 
by~\cite{horst} 
\begin{eqnarray}
 {\cal E}(\rho) &=& 
     \frac{2}{\pi^2}\int_{0}^{k_{F}} dk \,
     k^{2}\sqrt{k^2+M^{*2}} \nonumber \\
                &+& 
     \frac{1}{2}\left(\frac{m_{\rm s}^{2}}{g_{\rm s}^{2}}\right)
     \Phi_{0}^{2} + \frac{\kappa}{6}\Phi_{0}^{3} +
     \frac{\lambda}{24}\Phi_{0}^{4} \nonumber \\
                &+& 
     \frac{1}{2}\left(\frac{m_{\rm v}^{2}}{g_{\rm v}^{2}}\right)
      W_{0}^{2} + \frac{\zeta}{8}W_{0}^{4} \;.
 \label{EDensity}
\end{eqnarray}
Note that the following definitions have been introduced:
$\Phi_{0}=g_{\rm s}\phi_{0}$ and $W_{0}=g_{\rm v}V_{0}$.
The equation of state for symmetric nuclear matter is 
displayed on the left panel of Fig~\ref{Fig1}. 
As advertised $\zeta$, and to a lesser extent $M^{*}$, 
are responsible for a softening of the equation of state 
at high density.

However, in order to compute the equation of state for neutron-rich
matter one must supplement the equation of state for symmetric nuclear
matter with the symmetry energy. The symmetry energy, a
positive-definite quantity, is imposed as a penalty on the system for
upsetting the $N\!=\!Z$ balance. It is given by
\begin{equation}
  a_{\rm sym}(\rho) = \frac{k_{F}^{2}}{6E_{F}^{*}} 
          + \frac{g_{\rho}^{2}}{12\pi^{2}}
            \frac{k_{F}^{3}}{m_{\rho}^{*2}} \;,
 \label{SymmE}
\end{equation}   
where $E_{F}^{*}\!=\!\sqrt{k_{\rm F}^{2}+M^{*2}}$
and the ``effective'' rho-meson mass has been defined as
\begin{equation}
  m_{\rho}^{*2} = m_{\rho}^{2} + 2g_{\rho}^{2}
  \Big(\Lambda_{\rm s} \Phi_{0}^2 +
       \Lambda_{\rm v} W_{0}^{2} \Big) \;.
 \label{Mrho2}
\end{equation}     

In this manner, the equation of state of neutron-rich matter may 
be written as
\begin{equation}
  \frac{E}{A}(\rho,t) = \frac{{\cal E}(\rho)}{\rho} 
	              + t^{2} a_{\rm sym}(\rho) 
                      + {\cal O}(t^{4}) \;,
 \label{EOS}
\end{equation}     
where the neutron excess has been defined as
\begin{equation}
    t  \equiv \frac{\rho_{n} - \rho_{p}}
                   {\rho_{n} + \rho_{p}} \;.
 \label{tDef}
\end{equation}     

The symmetry energy is given as a sum of two contributions. The first
term in Eq.~(\ref{SymmE}) represents the increase in the kinetic
energy of the system due to the displacement of the Fermi levels of
the two species (neutrons and protons). This contribution has been
fixed by the properties of symmetric nuclear matter as it only depends
on the nucleon effective mass $M^{*}$. By itself, it leads to an
unrealistically low value for the symmetry energy; for example, at
saturation density this contribution yields $\sim\!15$~MeV, rather
than the most realistic value of $\sim\!37$~MeV. The second
contribution is due to the coupling of the rho meson to an
isovector-vector current that no longer vanishes in the $N\!\ne\!Z$
system. It is by adjusting the strength of the $NN\rho$ coupling
constant that one can now fit the empirical value of the symmetry
energy at saturation density. However, the symmetry energy at
saturation is not well constrained experimentally. Yet an average of
the symmetry energy at saturation density and the surface symmetry
energy is constrained by the binding energy of nuclei.  Thus, the
following prescription is adopted: the value of the $NN\rho$ coupling
constant is adjusted so that all parameter sets have a symmetry energy
of $25.7$~MeV at $k_F\!=\!1.15$~fm$^{-1}$~\cite{Ho01}. Following this
prescription the symmetry energy at saturation density is predicted to
be $37.3$, $36.6$, and $36.3$~MeV in the NL3, S271, and Z271 models,
respectively (for $\Lambda_{\rm s}\!=\!\Lambda_{\rm v}\!=\!0$).

The simplicity of the symmetry energy is remarkable indeed. The
contribution from the nucleon kinetic energy displays a weak model
dependence through the effective nucleon mass and this dependence
disappears in the high-density limit. The second term in
Eq.~(\ref{SymmE}) is also weakly model dependent, at least in the
$\Lambda_{\rm s}\!=\!\Lambda_{\rm v}\!=\!0$ limit. The reason is
simple: models constrained to reproduce the symmetry energy of nuclear
matter at some average density, while maintaining the effective
nucleon mass within the ``acceptable'' range of $0.6\!\le M^{*}/M\!\le
0.8$, yield values for the $NN\rho$ coupling constant within 15\% of
each other. The weak model dependence of the symmetry energy can be
observed in the right panel of Fig.~\ref{Fig1}. Note that the
high-density behavior of the symmetry energy is given by
\begin{equation}
  a_{\rm sym}(\rho) 
     \mathop{\longrightarrow}_{k_{F}\rightarrow\infty}
     \frac{g_{\rho}^{2}}{12\pi^{2}}
     \frac{k_{F}^{3}}{m_{\rho}^{2}} \;.
 \label{LimSymmE}
\end{equation}   

As they stand now, the models lack enough leverage to significantly
modify the symmetry energy. In order to remedy this deficiency one
must rely on the two nonlinear couplings between the isoscalar and
isovector mesons ($\Lambda_{\rm s}$ and $\Lambda_{\rm v}$). These 
couplings change the density dependence of the symmetry energy by 
modifying the rho-meson mass as is indicated in Eq.~(\ref{Mrho2}). 
For example, for $\Lambda_{\rm v}\!\ne\!0$ the high-density behavior 
of the omega-meson field becomes~\cite{horst}
\begin{equation}  
 W_{0} \mathop{\longrightarrow}_{k_{F}\rightarrow\infty}
       \cases{ 
\left(\displaystyle{\frac{g_{\rm v}^{2}}{m_{\rm v}^2}}\right)\!\rho
               & if $\zeta=0 \;;$ \cr
\left(\displaystyle{\frac{6\rho}{\zeta}}\right)^{1/3} 
               & if $\zeta \ne 0 \;.$ }
 \label{LimW0}
\end{equation}
In either case ($\zeta\!=\!0$ or $\zeta\!\ne\!0$) the modifications
are significant enough to change the qualitative behavior of the
symmetry energy; the symmetry energy now grows linearly with $k_{F}
\propto \rho^{1/3}$ rather than as $k_{F}^{3}$. This change in the
qualitative behavior of the symmetry energy can be seen in the right
panel of Fig.~\ref{Fig1}. Note that with $\Lambda_{\rm s} \!\ne\!0$ or
$\Lambda_{\rm v}\!\ne\!0$ an adjustment of the $NN\rho$ coupling
constant is necessary to maintain the symmetry energy unchanged from
its fixed value of $25.7$~MeV at $k_F\!=\!1.15$~fm$^{-1}$. Further,
the inclusion of these nonlinear terms does not affect the properties
of symmetric nuclear matter as ${\bf b}_\mu\!\equiv\!0$ in the
$N\!=\!Z$ limit.

\begin{figure}[h]
 \leavevmode\centering\psfig{file=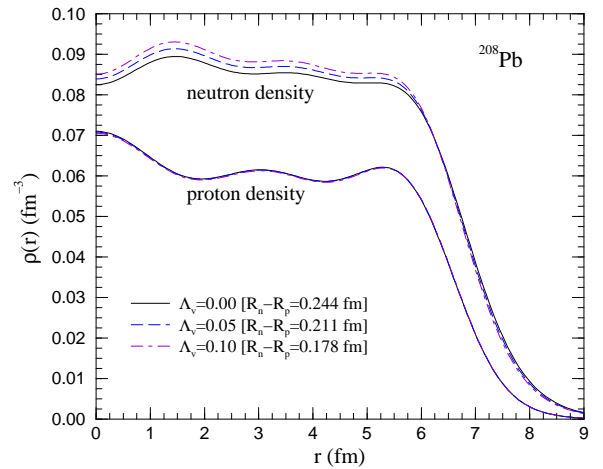,width=3.0in}
 \caption{Neutron and proton densities in $^{208}$Pb 
          in the Z271v model for three different values 
          of the nonlinear $\omega$-$\rho$ coupling 
          $\Lambda_{\rm v}$. In all cases the 
     	  root-mean-square charge radius is predicted 
         to be $R_{\rm ch}=5.51$~fm.}
 \label{Fig2}
\end{figure}

The procedure described above is robust in another important way.
While our goal is to induce changes in the neutron radius of
$^{208}$Pb through a modification of the symmetry energy, we want to
do so without sacrificing the success of the models in describing the
binding energy and charge radius of $^{208}$Pb, both of them well
known experimentally (${\rm B.E.}\!=\!7.868$~MeV and 
$R_{\rm ch}\!=\!5.51$~fm)~\cite{audi,sick,devries}. That this is 
possible may be seen in Fig.~\ref{Fig2}.  In this figure the neutron 
and proton ground-state densities have been computed in the Z271 model 
for three different values of the nonlinear $\omega$-$\rho$ coupling 
$\Lambda_{\rm v}$. While the softening of the symmetry energy has 
reduced the neutron skin of $^{208}$Pb appreciably, the charge radius 
has changed by less than $0.005$~fm.

\begin{figure}[h]
\leavevmode\centering\psfig{file=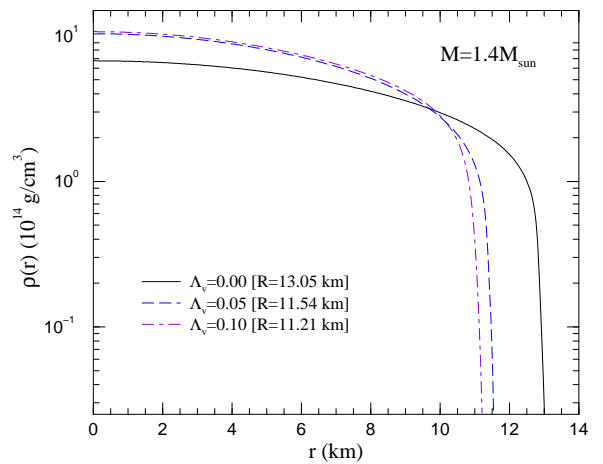,width=3.0in}
\caption{Density profile for a $M\!=\!1.4 M_{\odot}$ 
	 neutron star in the Z271v model for three 
	 different values of the nonlinear 
	 $\omega$-$\rho$ coupling $\Lambda_{\rm v}$.}
 \label{Fig3}
\end{figure}

In Fig.~\ref{Fig3} we show the equivalent plot but for an
object 55 orders of magnitude heavier than $^{208}$Pb: a 
1.4 solar-mass neutron star. The density profile of such 
a neutron star correlates nicely with the neutron skin of 
$^{208}$Pb. Models with a softer symmetry energy tolerate 
regions of large central densities thereby generating 
smaller radii.

\begin{figure}[h]
\leavevmode\centering\psfig{file=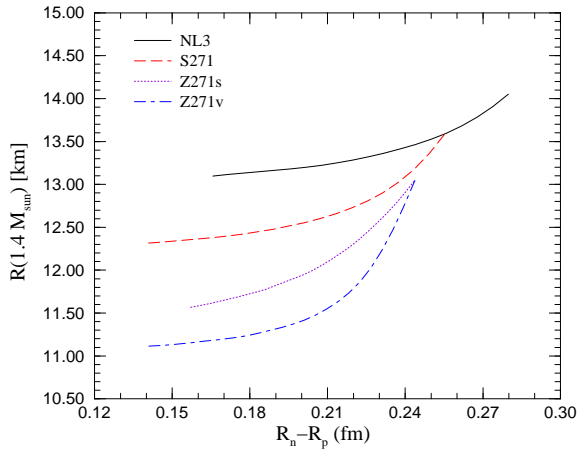,width=3.0in}
\caption{Radius of a $M\!=\!1.4 M_{\odot}$ neutron star  
         as a function of the neutron-minus-proton radius 
	 in $^{208}$Pb for the four parameter sets 
	 described in the text.}
 \label{Fig4}
\end{figure}

Finally, the radius $R$ of a 1.4 solar-mass neutron star as a function
of the neutron skin $R_n\!-\!R_p$ in $^{208}$Pb is displayed in
Fig.~\ref{Fig4} for the various models described in the text. All
neutron-star radii were computed using the Oppenheimer-Volkoff
equations for neutron-rich matter in beta equilibrium. Since this
figure was generated using an equation of state for uniform matter, it
may contain small errors due to an inappropriate treatment of the
surface region. Note that whereas the Z271 model has been extended to
include both $\Lambda_{\rm s}\!\ne\!0$ (Z271s) and 
$\Lambda_{\rm v}\!\ne\!0$ (Z271v), the other two models have the 
$\sigma$-$\rho$ coupling fixed at $\Lambda_{\rm s}\!=\!0$.  The strong
correlation between the neutron-star radius and the neutron skin in
$^{208}$Pb is evident: for a given parameter set $R$ increases with
$R_n\!-\!R_p$. However, as one modifies the parameter set to increase
$M^*$ or the $\omega$-meson self-coupling $\zeta$, the equation of
state becomes softer so the pressure decreases at high density. As a
result, the radius of the star becomes smaller for fixed
$R_n\!-\!R_p$. For example, for a neutron skin of
$R_n\!-\!R_p\!=\!0.18$~fm, the radius of the star varies from
$R\!\simeq\!13$~km in the NL3 model all the way down to
$R\!\simeq\!11$~km in the Z271v model. Thus, we conclude that the
radius of a $1.4 M_{\odot}$ neutron star is not uniquely constrained
by a measurement of the neutron-skin thickness because $R_n\!-\!R_p$
depends only on the equation of state at or below saturation density
while $R$ is mostly sensitive to the equation of state at higher
densities.  Yet one may be able to combine separate measurements of
$R_n\!-\!R_p$ and $R$ to obtain considerable information about the
equation of state at low and high densities. For example, if
$R_n\!-\!R_p$ is relatively large while $R$ is small, this could
indicate a phase transition. A large $R_n\!-\!R_p$ implies that the
low-density equation of state is stiff while a small $R$ suggests a
soft high-density equation of state. The change from stiff to soft
could be accompanied by a phase transition.

In conclusion, relativistic effective field theories that reproduce a
variety of ground-state observables have been used to correlate the
radius of a 1.4 solar-mass neutron star to the neutron skin of
$^{208}$Pb. Nonlinear couplings between isoscalar and isovector mesons
have been introduced to modify the density dependence of the symmetry
energy. Models with a softer symmetry energy tolerate larger central
densities and produce systems with smaller radii. Thus an important
correlation is revealed: the smaller the skin-thickness of $^{208}$Pb
the smaller the size of the neutron star. Yet the radius of the
neutron star is not uniquely constrained by a measurement of the
neutron skin in $^{208}$Pb. This is because the $^{208}$Pb measurement
constraints the equation of state only for densities between the
transition density to non-uniform matter and saturation density. In
contrast, the radius of a $1.4 M_{\odot}$ neutron star is mostly
sensitive to the equation of state at high density. Yet together they
provide considerable information on the equation of state.  If these
combined measurements reveal a large value of the neutron skin
together with a small value of the star radius, this may provide
strong evidence in support of a phase transition in the interior of
the neutron star.


\smallskip
This work was supported in part by DOE grants DE-FG02-87ER40365 
and DE-FG05-92ER40750.

\end{document}